\documentclass[conference]{IEEEtran}

\IEEEoverridecommandlockouts

\usepackage{cite}
\usepackage{amsmath,amssymb,amsfonts}
\usepackage{algorithmic}
\usepackage{graphicx}
\usepackage{textcomp}
\usepackage{xcolor}
\def\BibTeX{{\rm B\kern-.05em{\sc i\kern-.025em b}\kern-.08em
    T\kern-.1667em\lower.7ex\hbox{E}\kern-.125emX}}

\graphicspath{{./}} 

\usepackage{subfig}

\makeatletter
\newcommand{\linebreakand}{%
  \end{@IEEEauthorhalign}
  \hfill\mbox{}\par
  \mbox{}\hfill\begin{@IEEEauthorhalign}
}
\makeatother

\begin{document}

\title{Full-Scale GPU-Accelerated Transient EM-Thermal-Mechanical Co-Simulation for Early-Stage Design of Advanced Packages
\thanks{This work was supported by Rapid-HI (Heterogeneous Integration) Design
Institute (an Elmore ECE Emerging Frontiers Center), a grant from NSF
under award CCF-2235414, a grant from SRC under award 2878.033, and a DARPA NGMM grant.

\vspace{0.5em} \noindent \copyright~2026 IEEE. Personal use of this material is permitted.  Permission from IEEE must be obtained for all other uses, in any current or future media, including reprinting/republishing this material for advertising or promotional purposes, creating new collective works, for resale or redistribution to servers or lists, or reuse of any copyrighted component of this work in other works.}
}

\author{\IEEEauthorblockN{Hongyang Liu}
\IEEEauthorblockA{\textit{School of Elect. \& Comp. Eng.} \\
\textit{Purdue University}\\
West Lafayette, USA \\
liu3572@purdue.edu}
\and
\IEEEauthorblockN{Tejas Kulkarni}
\IEEEauthorblockA{\textit{School of Mechanical Engineering} \\
\textit{Purdue University}\\
West Lafayette, USA \\
kulkar72@purdue.edu}
\linebreakand
\IEEEauthorblockN{Ganesh Subbarayan}
\IEEEauthorblockA{\textit{School of Mechanical Engineering} \\
\textit{Purdue University}\\
West Lafayette, USA \\
ganeshs@purdue.edu}
\and
\IEEEauthorblockN{Cheng-Kok Koh}
\IEEEauthorblockA{\textit{School of Elect. \& Comp. Eng.} \\
\textit{Purdue University}\\
West Lafayette, USA \\
chengkok@purdue.edu}
\and
\IEEEauthorblockN{Dan Jiao}
\IEEEauthorblockA{\textit{School of Elect. \& Comp. Eng.} \\
\textit{Purdue University}\\
West Lafayette, USA \\
djiao@purdue.edu}
}

\maketitle

\begin{abstract}
In the early-stage design of advanced electronic packages, designers face a critical trade-off between simulation fidelity and computational turnaround time. Conventional early-stage methodologies typically achieve speed by relying on steady-state assumptions and structural homogenization. While computationally efficient, these approximations fundamentally fail to capture dynamic thermal events and stress concentrations at fine-grained internal interfaces, effectively masking failure mechanisms driven by transient signal bursts. In this work, we present a GPU-accelerated transient coupled Electromagnetic-Thermal-Mechanical solver that resolves this bottleneck. The proposed solver enables full-scale, non-homogenized, time-domain simulation of large-scale packages with runtimes amenable for rapid design iteration. Simulation of a NEC SX-Aurora TSUBASA package demonstrates that the tool allows for the identification of signal-induced adiabatic stress that is typically invisible to steady-state and homogenized baselines. This capability brings sign-off level physics fidelity to the early design phase, facilitating the prevention of costly late-stage design failures and broader transient thermal performance degradation risks.
\end{abstract}

\begin{IEEEkeywords}
Advanced packaging, early-stage design, transient multiphysics co-simulation, explicit structural resolution, GPU acceleration. 
\end{IEEEkeywords}

\vspace{-0.2in}
\section{Introduction} \label{sec:introduction}
\IEEEPARstart{T}{he} design of advanced electronic packages, characterized by 2.5D/3D integration and heterogeneous chiplet architectures, is an iterative process where decisions made in the early pathfinding stage have the most significant impact on final product success. Ideally, designers would evaluate the detailed multiphysics performance of every design variation to identify potential failure points immediately. However, the structural complexity of modern heterogeneous assemblies, which feature intricate multi-scale interconnects and dense integration features, imposes a severe computational burden.

Currently, this burden forces a rigid trade-off between simulation speed and physics fidelity. To achieve the rapid turnaround times required for design exploration, engineers are often compelled to rely on simplified modeling techniques. A pervasive approximation is the reliance on idealized or fictitious static power maps, rather than solving the actual electromagnetic physics \cite{Ref:StaticPowerMap}. This approach inherently neglects the complex, non-uniform current distributions driven by real high-speed signals and the resulting dynamic thermal response. Furthermore, to reduce mesh complexity, analysts frequently employ structural homogenization where intricate geometric details are replaced by effective material blocks \cite{Ref:Homogenization}.  

While these approaches allow for quick estimation of global metrics such as package-level warpage, they create a critical fidelity gap in the early design phase \cite{Ref:FidelityGap}. Static power maps average out the rapid, adiabatic heating events caused by transient signal bursts, while homogenization smears out the resulting stress concentrations at specific internal material boundaries. Consequently, critical failure mechanisms are often missed during pathfinding and are only discovered during expensive final verification stages \cite{Ref:LateStageFailure}. There is a critical need for simulation tools that can capture these transient multiphysics interactions on full-scale geometry while maintaining a computational cost that is manageable for iterative design.

In this work, we bridge this gap by developing a GPU-accelerated transient coupled Electromagnetic-Thermal-Mechanical solver tailored for early-stage design exploration. By leveraging advanced multiphysics simulation algorithms as well as optimized numerical libraries with custom high-performance CUDA kernels, the proposed solver achieves the computational throughput necessary to iterate on full-scale designs without sacrificing physics fidelity. This capability enables designers to simulate full-wave electromagnetic propagation, the resulting transient thermal shock, and the consequent mechanical stress on explicitly resolved structures. We demonstrate the solver's efficacy on an advanced package adapted from a real-world NEC SX-Aurora TSUBASA design \cite{Ref:TSUBASA}. The results highlight its ability to reveal transient, signal-induced adiabatic stress risks that conventional methodologies obscure, while also providing a rigorous basis for the analysis of broader device-level thermal degradation issues.
\section{Full-Scale Transient Multiphysics Analysis} \label{sec:formulation}
This section details the co-simulation algorithm of the proposed multiphysics solver of advanced packages. 

\subsection{Transient Electromagnetic-Thermal Coupling}
The electromagnetic analysis is based on an advanced matrix-free time-domain method \cite{Ref:MF-EM-Thermal} in a non-uniform grid, which results in the following discretization of Maxwell's equations,
\begin{equation}
\label{equ:Maxwell_Wave}
\overline{\mathbf{D}}_\epsilon \frac{d^2 \mathbf{e}}{d t^2} + \frac{d}{d t}\left(\overline{\mathbf{D}}_\sigma(t) \mathbf{e}\right) + \overline{\mathbf{S}} \mathbf{e} = -\frac{d \mathbf{J}}{d t},
\end{equation}
whose numerical system is free of matrix solution. Here, $\mathbf{e}$ denotes the unknown electric field intensity vector, and $\mathbf{J}$ represents the external current density excitation. $\overline{\mathbf{D}}_\epsilon$ and $\overline{\mathbf{D}}_\sigma$ are diagonal matrices representing the permittivity and temperature-dependent conductivity, respectively. The matrix $\overline{\mathbf{S}}$ corresponds to the discretized curl-curl operator $\nabla \times \mu^{-1} \nabla \times$, with $\mu$ denoting permeability.

Applying an explicit central difference scheme for time discretization, we derive the update equation. Letting $n$ denote the current time step, the system update from $n$ to $n+1$ is given by:
\begin{equation}
\label{equ:Maxwell_Update}
\overline{\mathbf{M}}\mathbf{e}^{n+1} = \overline{\mathbf{N}}\mathbf{e}^n - \overline{\mathbf{K}}\mathbf{e}^{n-1} - \Delta t^2 \left(\frac{\partial \mathbf{J}}{\partial t}\right)^n,
\end{equation}
where $\Delta t$ is the time step constrained by the Courant-Friedrichs-Lewy (CFL) stability condition. The system matrices are defined as:
\begin{equation}
\begin{gathered}
\overline{\mathbf{M}} = \overline{\mathbf{D}}_\epsilon + \frac{\Delta t}{2} \overline{\mathbf{D}}_\sigma(T^n), \quad
\overline{\mathbf{N}} = 2 \overline{\mathbf{D}}_\epsilon - \Delta t^2 \overline{\mathbf{S}}, \\
\overline{\mathbf{K}} = \overline{\mathbf{D}}_\epsilon - \frac{\Delta t}{2} \overline{\mathbf{D}}_\sigma(T^n).
\end{gathered}
\end{equation}
Here, $\overline{\mathbf{D}}_\sigma(T^n)$ indicates that conductivity is updated at each time step based on the instantaneous temperature distribution. Since $\overline{\mathbf{M}}$ is diagonal, the only computational cost of \eqref{equ:Maxwell_Update} is sparse matrix-vector multiplications, which are suitable for GPU acceleration.

Simultaneously, the transient heat diffusion equation is solved on the same finite-difference grid to ensure direct field coupling without interpolation errors. The discretized thermal equation is:
\begin{equation}
\label{equ:Thermal_Gov}
\tilde{\rho} c_p \frac{\partial \mathbf{T}}{\partial t} + \overline{\mathbf{M}}_{kk} \mathbf{T} = \mathbf{P}_{\text{total}},
\end{equation}
where $\mathbf{T}$ is the temperature vector, $\tilde{\rho}$ is the material density, and $c_p$ is the specific heat capacity. $\overline{\mathbf{M}}_{kk}$ is the discretized scalar Laplacian operator $-\nabla \cdot k \nabla$. It is formulated as $\overline{\mathbf{M}}_{kk} = \overline{\mathbf{V}}_{0a}^T \overline{\mathbf{D}}_k \overline{\mathbf{V}}_{0}$, where $\overline{\mathbf{V}}_{0}$ represents the discretized negative gradient operator ($-\nabla$), $\overline{\mathbf{V}}_{0a}^T$ represents the discretized divergence operator ($\nabla \cdot$), and $\overline{\mathbf{D}}_k$ is the thermal conductivity matrix. The term $\mathbf{P}_{\text{total}}$ is the Joule heating source vector derived from the electric field, $\mathbf{P}_{\text{total}} = \overline{\mathbf{D}}_\sigma \mathbf{e}^2$.

Using the Forward Euler method, the thermal time stepping is given by:
\begin{equation}
\label{equ:Thermal_Update}
\mathbf{T}^{n+1} = \mathbf{T}^n - \frac{\Delta t}{\tilde{\rho} c_p} \overline{\mathbf{M}}_{kk} \mathbf{T}^n + \frac{\Delta t}{\tilde{\rho} c_p} \mathbf{P}_{\text{total}}^{n+1},
\end{equation}
whose computation also involves sparse matrix-vector multiplications only, similar to the EM part. The two-way coupling is realized via the temperature dependence of the electrical conductivity:
\begin{equation}
\sigma(T) = \frac{\sigma_0}{1 + \alpha(T - T_0)},
\end{equation}
where $\sigma_0$ is the conductivity at reference temperature $T_0$, and $\alpha$ is the temperature coefficient of resistance.

A critical aspect of this formulation is the handling of the time-scale. In typical diffusion problems, the thermal time constant is orders of magnitude larger than the electromagnetic time step. However, this work focuses on transient adiabatic heating caused by fast signal spikes. In such scenarios, the temperature rise is nearly instantaneous and localized. Therefore, we strictly synchronize the thermal solver with the electromagnetic solver, using the same small $\Delta t$.

\subsection{Thermo-Mechanical Analysis}
The mechanical response is modeled using the theory of linear thermoelasticity. We solve for the displacement field $\mathbf{q}$ that satisfies the equations of equilibrium under thermal loading. The governing system consists of the strain-displacement relation, the conservation of momentum, and the constitutive law:
\begin{equation}
\label{equ:Mech_System}
\begin{gathered}
\epsilon_{ij} = \frac{1}{2}\left(\frac{\partial u_i}{\partial x_j} + \frac{\partial u_j}{\partial x_i}\right), \quad \frac{\partial \sigma_{ij}}{\partial x_j} + f_i = 0, \\
\sigma_{ij} = C_{ijkl}\left(\epsilon_{kl} - \alpha_{L} \delta_{kl} \Delta T\right),
\end{gathered}
\end{equation}
where $\epsilon_{ij}$ and $\sigma_{ij}$ are the strain and stress tensors, $u_i$ are displacement components, $C_{ijkl}$ is the stiffness tensor, $\alpha_L$ is the coefficient of linear thermal expansion (CTE), and $f_i$ denotes body forces. In this specific packaging analysis, we assume no external mechanical body forces ($f_i=0$); the deformation is driven entirely by the internal thermal stress term arising from the transient temperature distribution $\Delta T$.

The domain is discretized using 8-node hexahedral (HEX8) elements with $2\times2\times2$ Gauss quadrature. The resulting finite element system is:
\begin{equation}
\label{equ:Mech_FEM}
\left(\int_V \overline{\mathbf{B}}^T \overline{\mathbf{E}}_c \overline{\mathbf{B}} \, dV\right) \mathbf{q} = \int_V \overline{\mathbf{B}}^T \overline{\mathbf{E}}_c \boldsymbol{\epsilon}_T \, dV,
\end{equation}
where $\mathbf{q}$ is the global vector of nodal displacements and $\overline{\mathbf{B}}$ is the strain-displacement matrix containing shape function derivatives. $\overline{\mathbf{E}}_c$ is the standard isotropic elasticity matrix defined by the Lamé parameters $\lambda$ and $\mu$, which are derived from Young's modulus $E$ and Poisson's ratio $\nu$. The thermal strain vector is given by $\boldsymbol{\epsilon}_T = \alpha_L \Delta T [1, 1, 1, 0, 0, 0]^T$.

Finally, to assess reliability, the Von Mises stress $\sigma_{VM}$ is computed at the centroid of each element:
\begin{equation}
\begin{split}
\sigma_{VM} = \Big( \frac{1}{2}\big[ &(\sigma_x-\sigma_y)^2 + (\sigma_y-\sigma_z)^2 + (\sigma_z-\sigma_x)^2 \\
&+ 6(\tau_{xy}^2 + \tau_{yz}^2 + \tau_{zx}^2) \big] \Big)^{1/2}.
\end{split}
\end{equation}
This scalar metric provides a widely accepted criterion for predicting yield and structural failure in ductile materials used in electronic packaging.
\section{Simulation Setup} \label{sec:setup}
We applied the proposed multiphysics simulator to analyze a large-scale advanced package adapted from the NEC SX-Aurora TSUBASA architecture \cite{Ref:TSUBASA}. This section details the geometric configuration, material properties, and the transient excitation scheme.

\subsection{Structure and Material Configuration}
\begin{figure}[t]
\centering
\includegraphics[width=0.6\columnwidth]{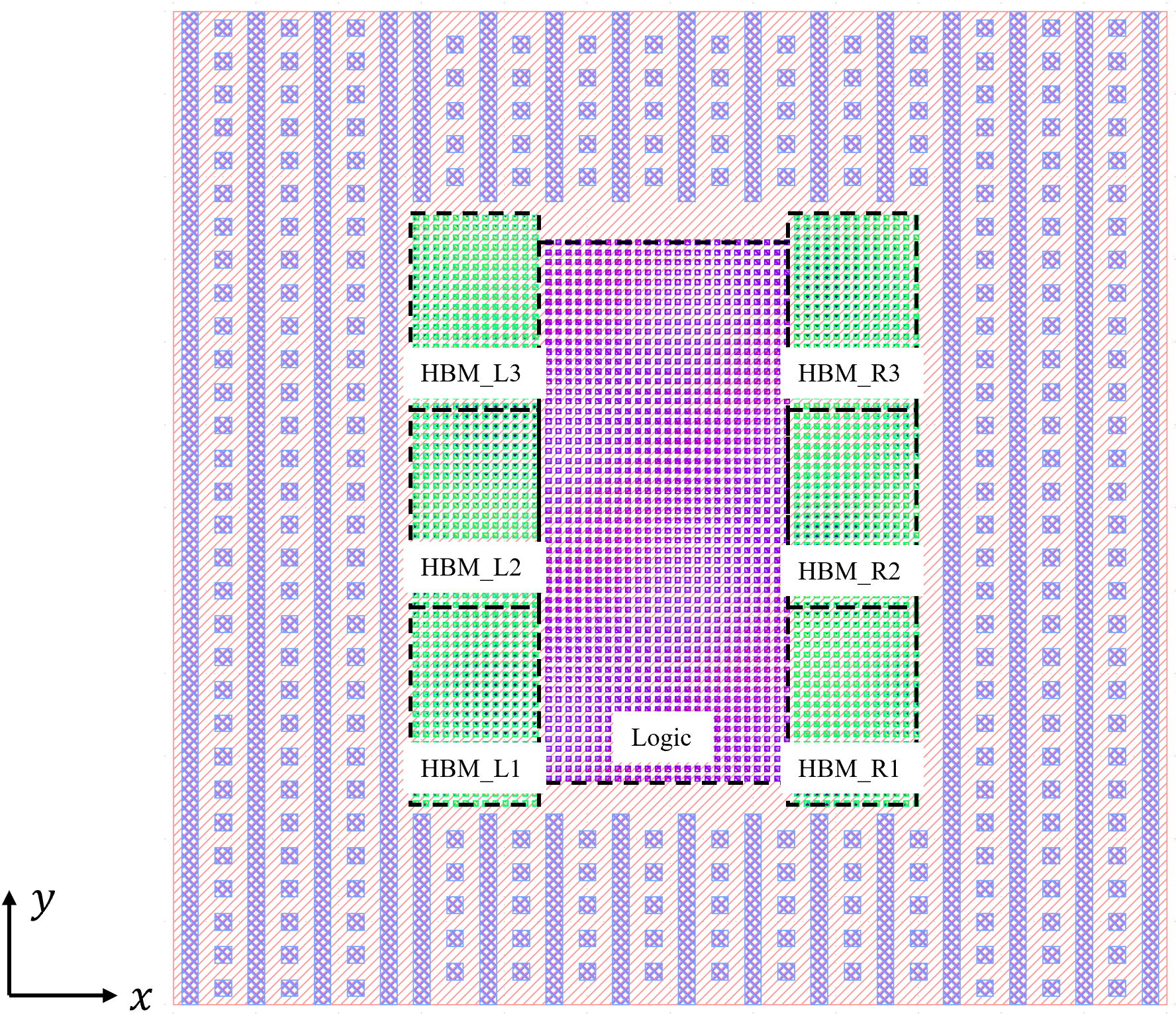}
\caption{GDS layout of the simulated advanced package structure showing the distribution of logic and HBM blocks.}
\vspace{-0.2in}
\label{fig:layout}
\end{figure}

The simulation domain spans $60 \times 60$ mm$^2$ in the lateral dimensions and has a total thickness of $2.80$ mm. The vertical stack comprises 26 unique layers transitioning from a laminated organic substrate (0.5--1.0 mm thick) to a silicon interposer (0.1--0.3 mm thick), and finally to a heterogeneous assembly of active logic slices and High Bandwidth Memory (HBM) stacks (0.015--0.165 mm thick). All interfaces are assumed to be perfectly bonded, with no slip or delamination. The structure features complex electrical details, including dense micro-bump arrays and differential routing. A continuous ground plane covers the bottom surface ($z=0$), and the layout is shown in Fig. \ref{fig:layout}. 

Crucially, this work avoids material homogenization. To accurately capture local stress concentrations at interfaces, we assign distinct linear elastic isotropic material properties to all geometric features. We utilize standard packaging values for Copper ($E=128$ GPa, $\alpha=17$ ppm), Solder ($E=50$ GPa, $\alpha=25$ ppm), Silicon ($E=130$ GPa, $\alpha=2.6$ ppm), as well as SiO$_2$, Polyimide, Underfill, Epoxy Mold, and BT Resin.

\subsection{Boundary Conditions and Excitation Scheme}
Appropriate boundary conditions are applied to mimic the operating environment. For the electromagnetic solver, the computational domain is truncated with Perfect Magnetic Conductor (PMC) boundaries to approximate an open system. For the thermal solver, an adiabatic condition is assumed at all boundaries. This is justified by the extremely short duration of the transient events, during which heat does not have sufficient time to diffuse to the package exterior or heat sinks. For the mechanical solver, we employ a kinematic mount (minimum constraint) configuration to prevent rigid body motion while allowing unconstrained thermal expansion. Specifically, the node at the geometric center of the bottom surface ($z=0$) is fixed in all three dimensions ($x, y, z$). A node on the right edge of the bottom surface (same $y$ coordinate as the center) is fixed in $y$ and $z$ but free in $x$, allowing lateral expansion. Finally, a node on the top edge of the bottom surface (same $x$ coordinate as the center) is fixed in $z$ but free in $x$ and $y$. This setup effectively anchors the package without inducing artificial stress.

The structure is subjected to a complex transient excitation scheme designed to simulate simultaneous power delivery noise, logic switching, and high-speed data transmission. The excitation signals are spatially distributed across the substrate, logic, and HBM regions. In the substrate, differential signaling is simulated using $x$-oriented ports with alternating polarities along the $y$-direction. In the HBM and logic bump arrays, a Ground-Signal-Ground (G-S-G) configuration is used with $x$-oriented differential feeding. This local current sinking mimics realistic return paths. Three distinct temporal waveforms are superimposed to create the total transient load: (1) a broadband ``Trigger'' event (20--60 ps) consisting of a Ricker wavelet ($t_0=40$ ps) applied to the substrate; (2) a background ``Logic Noise'' (0--300 ps) consisting of a continuous 5 GHz sine wave applied to the logic die; and (3) a high-speed ``Data Burst'' (80--230 ps) consisting of a 40 GHz carrier modulated by a 3-bit Gaussian pulse train applied to the HBM stacks.

To reflect manufacturing and operational variability, a deterministic random jitter is applied to each excitation port. The amplitude, pulse width, and temporal phase are varied by up to $\pm 20\%$, $\pm 10\%$, and $\pm 10$ ps, respectively. This ensures that the differential signals are not perfectly balanced, generating realistic common-mode noise and localized hotspots.
\vspace{-0.1in}
\section{Simulation Results} \label{sec:results}
\subsection{Implementation and Performance}
The proposed multiphysics simulation is implemented in C++ using the NVIDIA cuSPARSE library and custom high-performance CUDA kernels to handle the multiphysics coupling. The transient coupled electromagnetic and thermal updates are handled explicitly. For the mechanical linear system, we utilize the NVIDIA AmgX library on an NVIDIA A100 80GB GPU, employing a Flexible GMRES (FGMRES) solver preconditioned by an Algebraic Multigrid (AMG) method. This configuration ensures robust convergence for the ill-conditioned elasticity matrices characteristic of heterogeneous packaging structures.

The test structure is discretized into a mesh of $376 \times 408 \times 45$ cells in the $x, y,$ and $z$ directions, respectively. This resolution results in large degrees of freedom (DOF) totaling 21,088,285 for the electromagnetic solver, 7,092,878 for the thermal solver, and 21,278,634 for the mechanical solver. We simulate a transient observation window of 300 ps with a time step of $\Delta t = 20$ fs. The total wall-clock time for the coupled time marching is approximately 79.71 seconds. Each static mechanical solve, triggered at peak temperature snapshots, requires approximately 62 seconds. The resulting total execution time is on the order of minutes, providing a computational throughput commensurate with the rapid turnaround requirements of early-stage design exploration.

\subsection{Transient Fields and Reliability Analysis}
While the solver computes full 3D fields for the entire structure, we select the interface between the substrate and the interposer ($z=1.6$ mm, Layer 2 C4 Bump layer) as a representative example to demonstrate the transient behavior.

\begin{figure}[!t]
\centering
\subfloat{
\includegraphics[width=0.48\columnwidth]{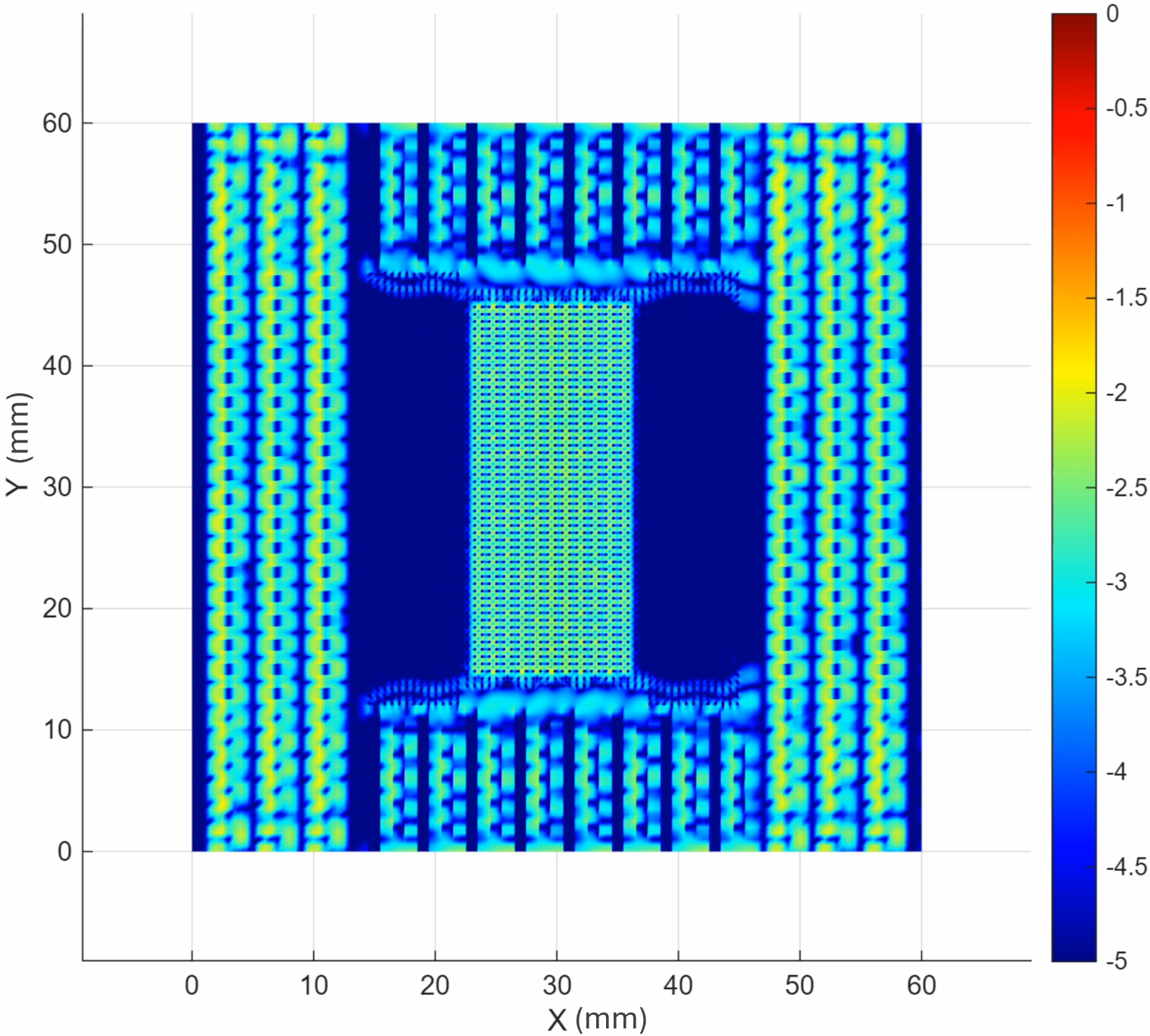}
\label{fig:power_t1}}
\subfloat{
\includegraphics[width=0.48\columnwidth]{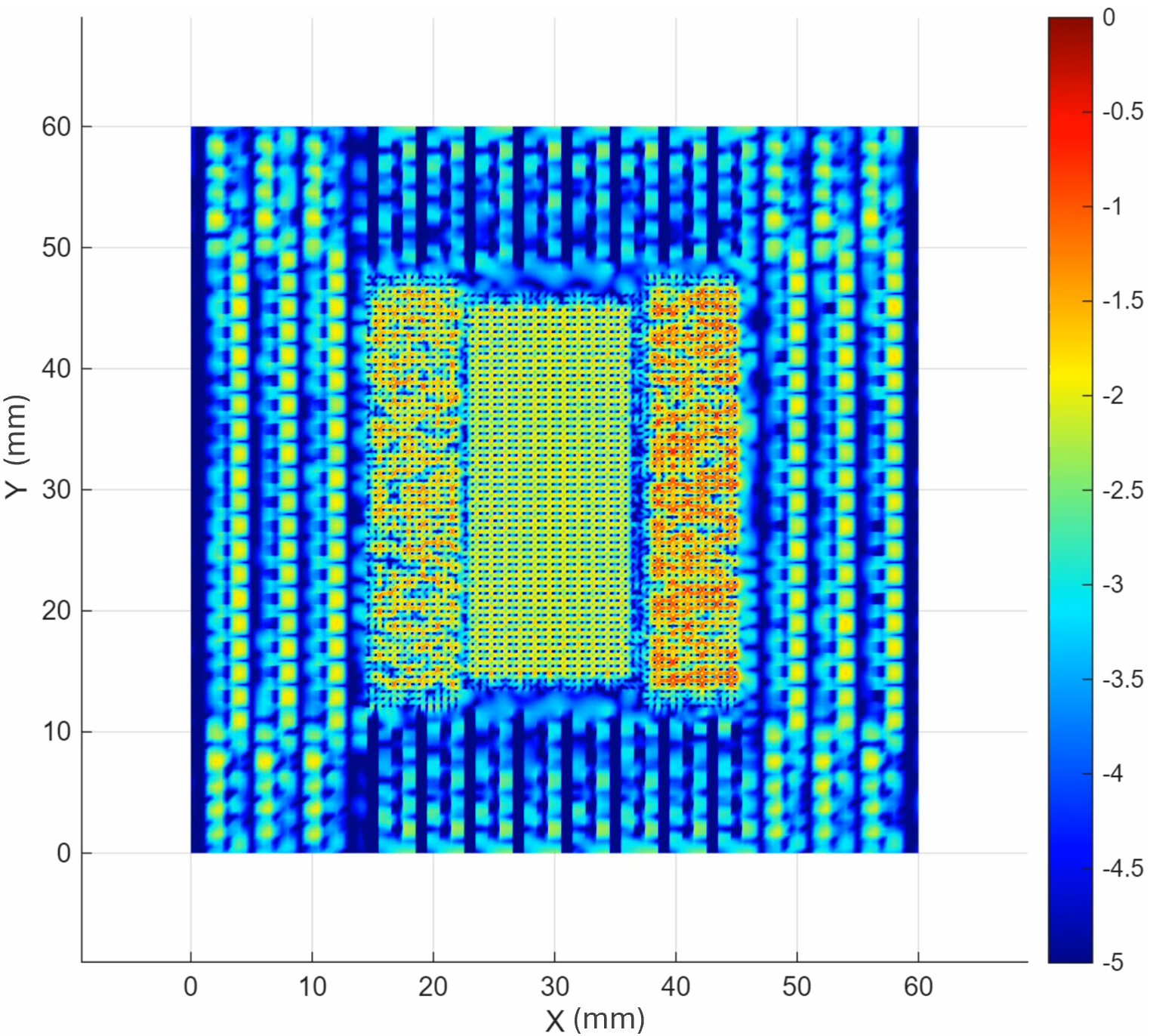}
\label{fig:power_t2}}
\caption{Normalized transient Joule power distribution at $z=1.6$ mm (Interposer C4 layer) at (a) $t=79$ ps and (b) $t=166$ ps. 0 dB corresponds to the peak value.}
\label{fig:transient_power}
\vspace{-0.2in}
\end{figure}

First, we examine the instantaneous Joule heating power density. Fig. \ref{fig:transient_power} displays the normalized power distribution (in dB, with a -5 dB floor). These snapshots capture the dynamic nature of the excitation, illustrating how energy injection shifts spatially from the substrate vias to the logic and HBM bumps, acting as localized, time-varying heat sources.

\begin{figure}[!t]
\centering
\includegraphics[width=0.6\columnwidth]{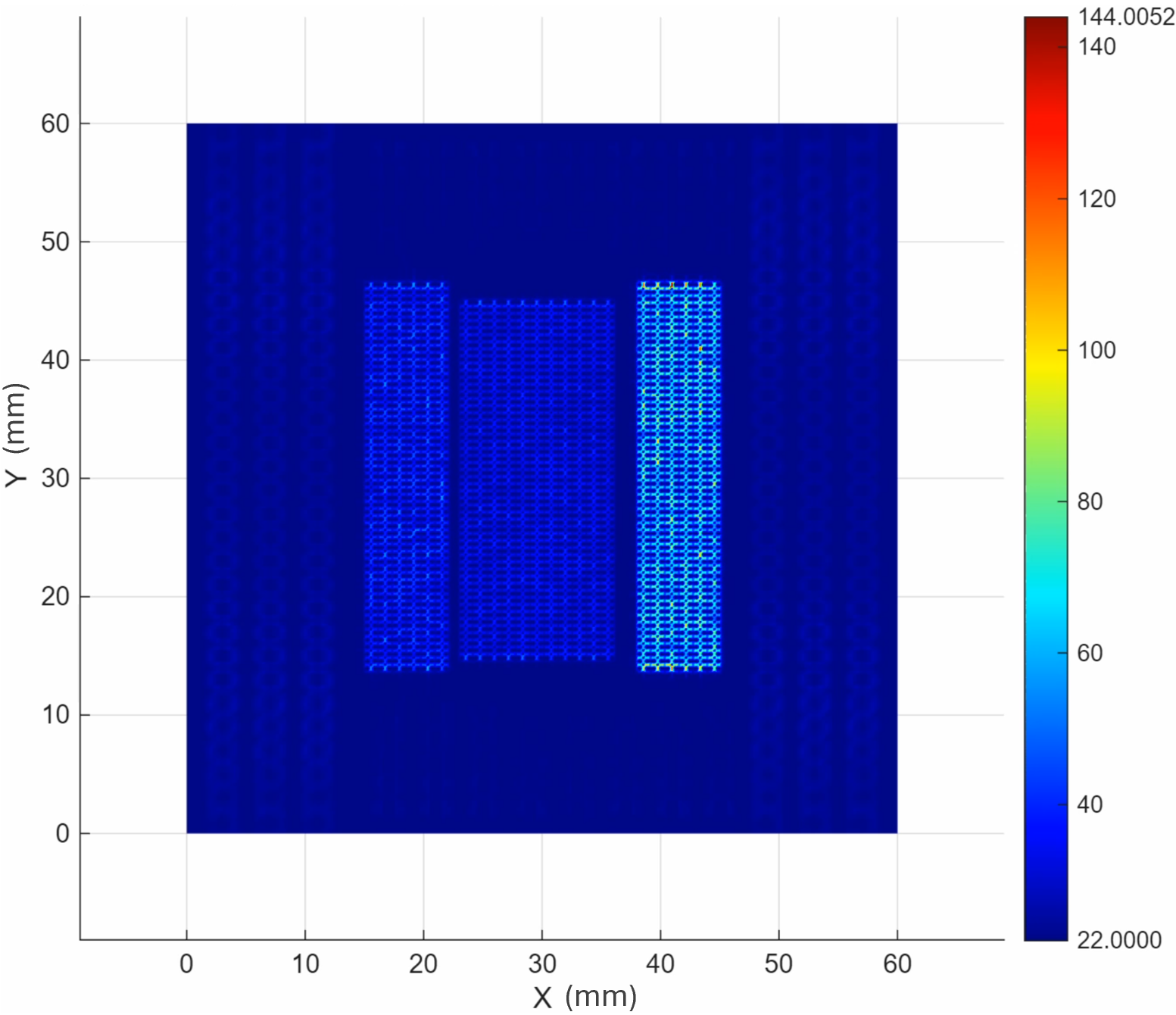}
\caption{Transient temperature distribution in Celsius at $z=1.6$ mm at $t=300$ ps. The distribution highlights localized heating missed by steady-state approximations.}
\label{fig:temp_map}
\vspace{-0.3in}
\end{figure}

The cumulative effect of this transient power is observed in the temperature distribution. Fig. \ref{fig:temp_map} shows the temperature field at the end of the pulse window ($t=300$ ps). Unlike steady-state analyses which produce smooth, diffused gradients, this result reveals sharp, adiabatic temperature spikes localized around the signal traces and micro-bumps. This confirms that significant thermal gradients can develop before heat diffusion equilibrates the system, creating a thermal shock regime that is distinct from the slow heating cycles typically modeled in package reliability.

\begin{figure}[!t]
\centering
\subfloat{
\includegraphics[width=0.48\columnwidth]{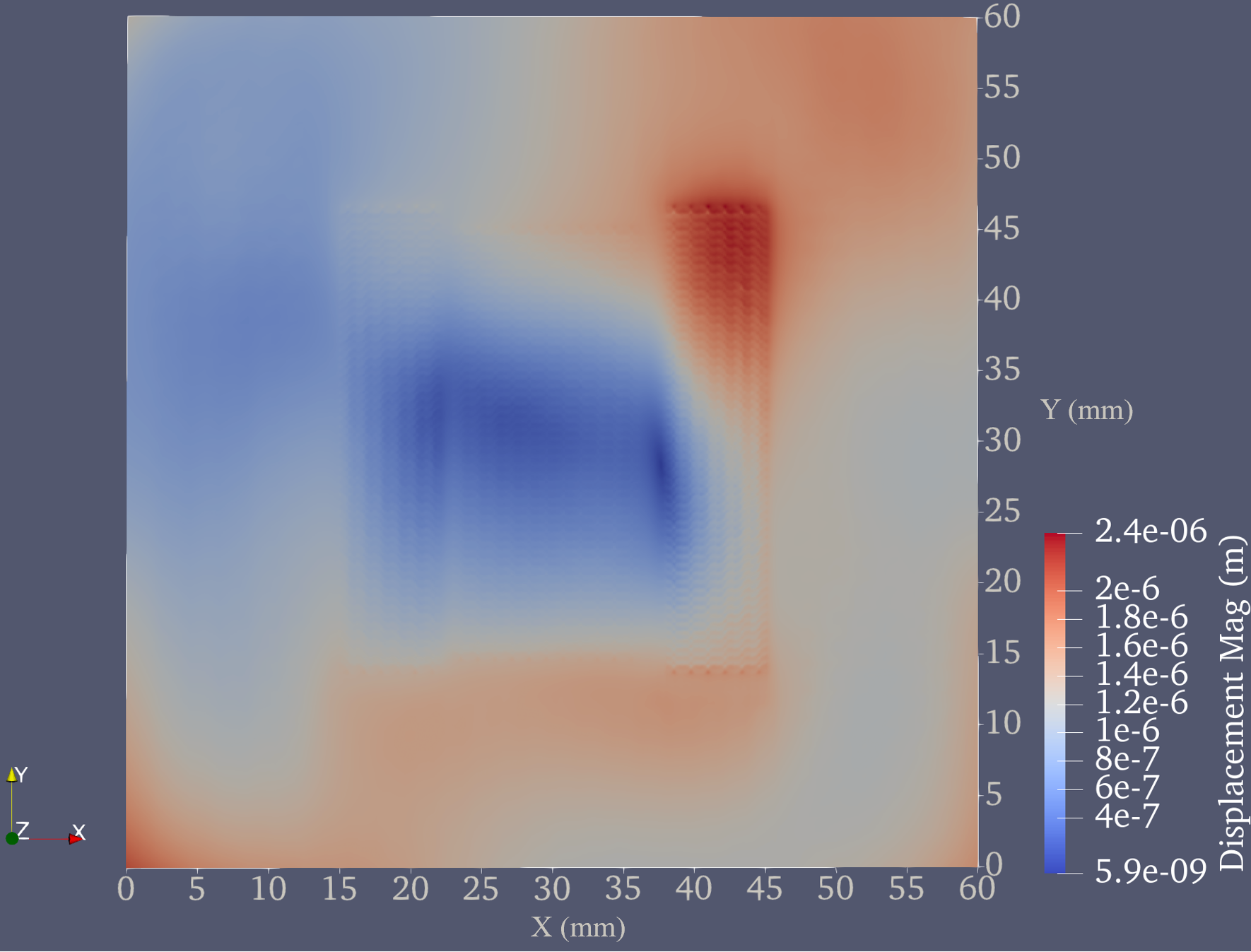}
\label{fig:displacement}}
\subfloat{
\includegraphics[width=0.5\columnwidth]{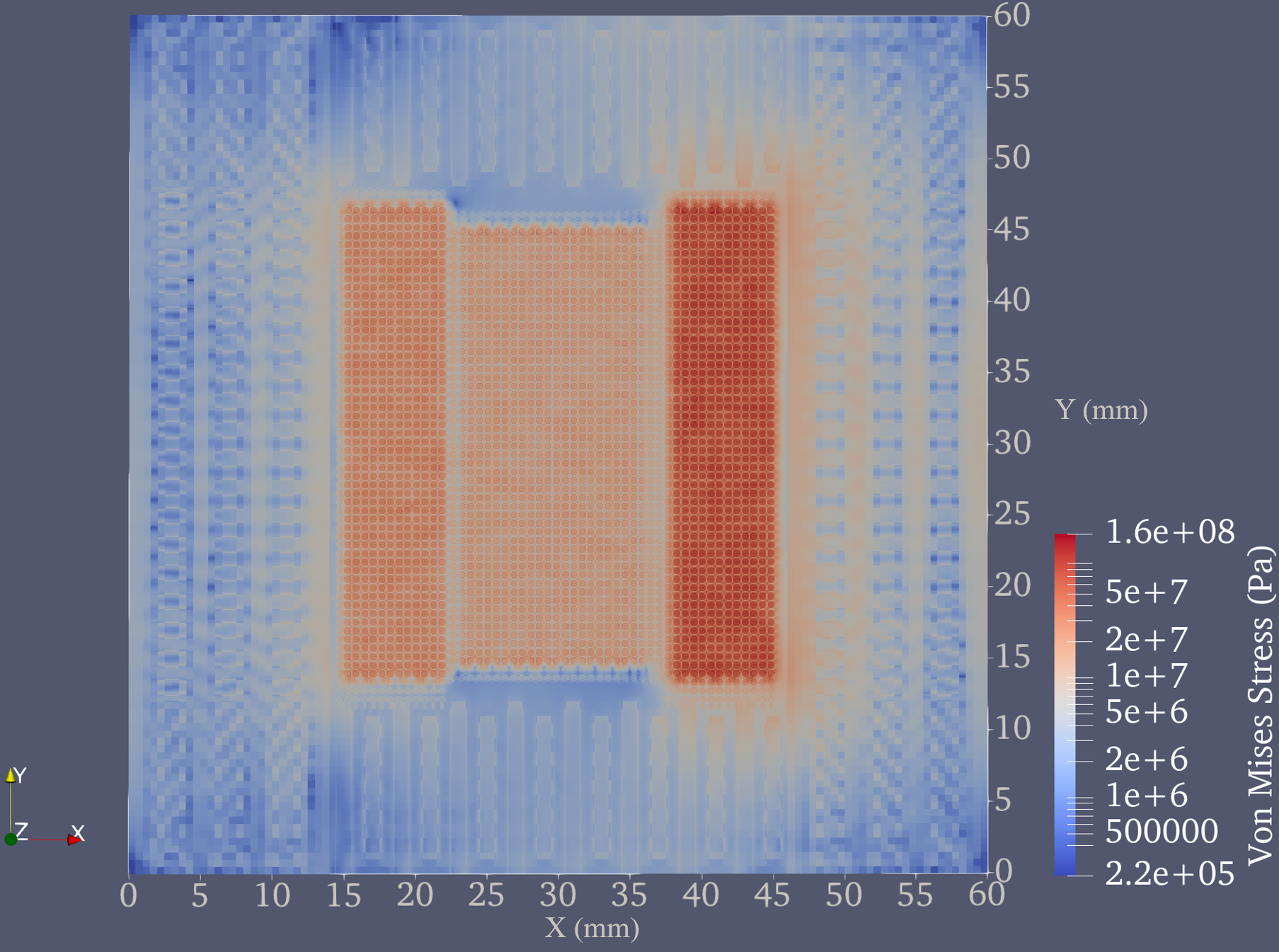}
\label{fig:stress}}
\caption{Mechanical response at $z=1.6$ mm derived from the transient thermal load: (a) Displacement magnitude, and (b) Von Mises stress distribution.}
\vspace{-0.2in}
\label{fig:mech_response}
\end{figure}

Finally, the mechanical response is computed based on the peak temperature profile. Fig. \ref{fig:mech_response} presents the displacement magnitude and the Von Mises stress distribution. The observed temperature spike drives uneven planar deformation at the substrate-interposer interface. Variations in displacement along the substrate plane introduce shear and out-of-plane loading, which is reflected in the stress distribution. A significant non-uniformity is observed, with high-stress concentrations appearing at the interfaces of materials with mismatched CTEs (e.g., between the copper bumps and the surrounding underfill/dielectric). These localized stress concentrations, occurring at the sub-bump scale, would be completely averaged out by standard homogenization techniques. By resolving the interface explicitly, the solver identifies critical indicators of delamination risk that bulk-property models obscure. This demonstrates the tool's capacity to reveal sign-off level reliability hazards during the early design phase, preventing defects that would otherwise propagate to later verification stages.
\vspace{-0.05in}

\end{document}